\documentclass[twocolumn]{autart}    
\pdfoutput=1
\usepackage{amsmath,amssymb,amsfonts}
\usepackage{graphicx}
\usepackage{multirow}
\usepackage{IEEEtrantools}
\usepackage[T1]{fontenc}
\usepackage[dvipsnames]{xcolor}
\usepackage{wrapfig}
\usepackage{natbib}
\usepackage[absolute,overlay]{textpos}

\newcommand{\diag}{\mathop{\mathrm{diag}}}
\newcommand{\row}{\mathop{\mathrm{row}}}
\newcommand{\trace}{\mathop{\mathrm{tr}}}

\begin{document}

\begin{frontmatter}
	
	\title{Data-Based Control of Continuous-Time Linear Systems with Performance Specifications\thanksref{footnoteinfo}} 
	
	\thanks[footnoteinfo]{This paper was not presented at any IFAC meeting. Corresponding author V.~G.~Lopez. Tel. +49 511 762 5189.}
	
	\author{Victor G. Lopez}\ead{lopez@irt.uni-hannover.de},    
	\author{Matthias A. M\"uller}\ead{mueller@irt.uni-hannover.de}
	\address{Leibniz University Hannover, Institute of Automatic Control, 30167 Hannover, Germany}  

	\begin{abstract}                          
		The design of direct data-based controllers has become a fundamental part of control theory research in the last few years. In this paper, we consider three classes of data-based state feedback control problems for linear systems. These control problems are such that, besides stabilization, some additional performance requirements must be satisfied. First, we formulate and solve a trajectory-reference control problem, on which desired closed-loop trajectories are known and a controller that allows the system to closely follow those trajectories is computed. Then, the solution of the LQR problem for continuous-time systems is presented. Finally, we introduce a data-based variant of a robust pole-placement procedure, where the precise position of the desired closed-loop poles is known. The applicability of the proposed methods is tested using numerical simulations.
	\end{abstract}

	
\end{frontmatter}

\begin{textblock*}{19cm}(2.1cm,26.8cm) 
	\footnotesize \copyright 2026. Manuscript available under the CC-BY-NC-ND 4.0 license https://creativecommons.org/licenses/by-nc-nd/4.0/ \\
	Article published in Automatica. DOI 10.1016/j.automatica.2026.113002
\end{textblock*}

\section{Introduction}
\label{sec:introduction}
Designing stabilizing controllers directly from measured data, without resorting to an explicit model identification step, has been the focus of plenty of the recent research in control theory. In the case of discrete-time (DT) systems, the result known as Willems' fundamental lemma \citep{WillemsRapMarDe2005} has been the basis of much of the recent progress in this field. For an overview of many of these developments, see \cite{MarkovskyDor2021} and the references therein. Some works have addressed the control design for continuous-time (CT) systems, as in \cite{MarkovskyRa2008,BerberichWilHerAll2021,BisoffiPerTes2022,EisingCor2023,RapisardaWaaCam2023}. The main goal of many of these works is to determine stabilizing controllers, without concerns about other closed-loop system characteristics. However, some results have been obtained for data-based control with performance specifications. In the following, we focus the discussion on works that aim to determine a state feedback gain, and omit other classes of controllers with performance guarantees as, for example, predictive control \citep{CoulsonLygDoe2019,BerberichKoeMueAll2021}.

A data-based solution to the discrete-time linear quadratic regulator (LQR) problem based on Willems' lemma was proposed in \cite{DePersisTes2020}. This result relies on the solution of an optimization problem with linear matrix inequality (LMI) constraints. Other data-based solutions of the discrete-time LQR problem using LMIs have been studied in \cite{vanWaardeEisTreCam2020,DorflerTesPer2023}. Different from the use of LMIs, methods based on reinforcement learning have also been used to determine model-free optimal controllers for discrete-time \citep{KiumarsiLewJia2017,LopezAlsMue2021} and continuous-time systems \citep{JiangJia2012,LopezMulCDC2023}. Data-based pole placement procedures have also been investigated \citep{Bianchin2023,BisoffiPerTes2023}, providing a different type of closed-loop performance specification. The method in \cite{BisoffiPerTes2023} has robustness properties against noise in the data, but this result does not consider the problem of \emph{exact} pole placement. Instead, LMIs are used to define desired regions of the complex plane for the closed-loop poles. Although model-based exact pole placement methods with robustness against model uncertainties are known \citep{SchmidPanNgu2014,Varga2000}, data-based variants have not yet been reported. In model-reference control problems, a feedback controller is designed such that the closed-loop system emulates the behavior of a desired reference model. An example of a data-based model-reference method is \emph{virtual reference feedback tuning} \citep{CampiLecSav2002}. A different approach was followed in \cite{BreschiPerForTes2021}, where LMIs are leveraged to obtain a discrete-time model-reference controller.

In this paper, we formulate and solve three different classes of data-based control problems for CT systems with performance specifications. For the first problem, we introduce a \emph{trajectory-reference} controller. Here, instead of a reference model, only data about desired closed-loop state trajectories are available. This is the case, for example, when state data from an expert system are available (e.g., a human driving a vehicle \citep{HuangWuLv2023}), and a controller must be designed to imitate such behavior. Although the trajectory-reference control problem is related to the model-reference problem, the former can be solved directly when only data about the desired trajectories are known. The second class of considered controllers is the data-based LQR problem for \emph{continuous-time} systems. Finally, for the third class of controllers, we present a robust exact pole placement algorithm that corresponds to a data-based version of the method described in \cite{SchmidPanNgu2014}. 
 
The contributions of this paper are the following. First, we present a set of data-based conditions for stability of CT systems with noisy data and establish the connection between these conditions and the data-based system representation in \cite{LopezMuCDC2022}. We also show that the obtained conditions recover various known stability results in the literature. Then, using the developed control framework, we formulate and solve the three continuous-time data-based control problems described above. To the best of our knowledge, the data-based formulation of the trajectory-reference problem is novel in the data-based control literature. The solution of the data-based LQR problem has been presented for DT systems, but not for the CT case. The proposed robust pole placement algorithm allows \textit{exact} pole placement (unlike \cite{BisoffiPerTes2023}) and provides robustness properties against noise (unlike \cite{Bianchin2023}).

In the following, Section~\ref{secprel} presents useful definitions to be used throughout the paper. In Section~\ref{seccond}, conditions for stabilization of continuous-time systems are obtained. Our solutions to three data-based problems are described in Section~\ref{sectr}. Section~\ref{secsim} presents numerical examples, and Section~\ref{secconc} concludes the paper.

\section{Preliminaries}
\label{secprel}

\subsection{Notation}
\label{secnot}

A symmetric, positive definite matrix is denoted as $P \succ 0$. The Frobenius norm of a matrix $M$ is $\| M \|_F$. For any matrix $M$, the notation $M_{(a:b)}$, for $a,b >0$, denotes the submatrix composed with the rows $a$ to $b$ of $M$. $\diag \{ M_i \}_{i=1}^q$ represents a block-diagonal matrix with diagonal block elements $M_i$, $i=1,\ldots,q$. Similarly, $M= \row \{ M_i \}_{i=1}^q$ denotes the block-row matrix defined by the horizontal concatenation $M := [M_1 \cdots M_q]$. The following definition will be of particular use in Section~\ref{secpp}. Consider a matrix $M = \row \{ M_i \}_{i=1}^q$ defined such that, if the block element $M_i$ is complex, then its complex conjugate $M_i^*$ is also a block element of $M$. The matrix $\mathcal{R}(M)$ is then defined as a matrix with the same dimensions as $M$, but with each pair of complex conjugate blocks $M_i$ and $M_i^*$ replaced by the real matrices $(1/2)(M_i + M_i^*)$ and $(1/2j)(M_i - M_i^*)$, respectively. A set $\Lambda$ is said to be self-conjugate, if $\lambda_i \in \Lambda$ implies $\lambda_i^* \in \Lambda$.

Throughout this paper we consider a given integer $N \in \mathbb{N}$, a positive scalar $T~\in~\mathbb{R}_+$, and continuous-time signals of the form $\xi : [0, NT] \rightarrow \mathbb{R}^\sigma$, with $[0, NT] \subset \mathbb{R}$.  Moreover, for any CT signal $\xi$, define the matrices
\begin{equation}
	\mathcal{H}_{\mathcal{T}}(\xi(t)) := \left[ \begin{array}{cccc}
		\xi(t) & \xi(t+T) & \cdots & \xi(t+(N-1)T) \end{array} \right],
	\label{hankrow}
\end{equation}
where we use $\mathcal{T}$ to denote the pair $\mathcal{T} := (N,T)$. Note that, different to the notation typically used in the DT case (compare, e.g., \cite{WillemsRapMarDe2005}), the subscript $\mathcal{T}$ in \eqref{hankrow} does \emph{not} correspond to the depth of the matrix.

\subsection{A continuous-time version of Willems' lemma}

Consider a continuous-time linear system of the form
\begin{equation}
	\dot x_{f}(t) = Ax_{f}(t) + Bu(t),
	\label{ctsys}
\end{equation}
where $x_{f} \in \mathbb{R}^n$ and $u \in \mathbb{R}^m$ are the state\footnote{The notation $x_{f}$ represents noise-free measurements. We use the notation $x$ for noisy measurements (see Section \ref{seccond}).} and input vectors of the system, respectively. The pair $(A,B)$ is assumed to be unknown but controllable. Consider the continuous-time input signal $u : [0, NT] \rightarrow \mathbb{R}^m$, as well as measurements of the state trajectory $x_{f}: [0, NT] \rightarrow \mathbb{R}^n$ and its derivative $\dot x_{f}: [0, NT] \rightarrow \mathbb{R}^n$ satisfying \eqref{ctsys}. The availability of the state derivative is a common assumption in the literature for data-based control of CT systems \citep{BisoffiPerTes2022,EisingCor2023,MillerSzn2023}. This derivative can be either measured or estimated using the available state data. As it is suggested in \cite[Appendix A]{PersisPosTes2022} and \cite{LopezMulCDC2023}, also in the present work the need for derivative information can be replaced by a data integration procedure. For simplicity, throughout the paper we assume the availability of state derivative trajectories. Defining the matrices $\mathcal{H}_{\mathcal{T}}(\cdot)$ as in \eqref{hankrow}, notice that (\ref{ctsys}) implies
\begin{equation}
	\mathcal{H}_{\mathcal{T}}(\dot x_{f}(t)) = A \mathcal{H}_{\mathcal{T}}(x_{f}(t)) + B \mathcal{H}_{\mathcal{T}}(u(t)).
	\label{hdyn}
\end{equation}

In \cite{LopezMuCDC2022}, a class of persistently exciting (PE) inputs for continuous-time systems is defined as follows.

\begin{defn}[Piecewise constant PE input]
	\label{defctpe}
	Consider a time interval $T > 0$ such that
	\begin{equation}
		T \neq \frac{2 \pi k}{| \mathrm{Im}(\lambda_i - \lambda_j)|}, \qquad \forall k \in \mathbb{Z},
		\label{assut}
	\end{equation}
	where $\lambda_i$ and $\lambda_j$ are any two eigenvalues of $A$ in (\ref{ctsys}), and $\mathrm{Im}(\cdot)$ is the imaginary part of a complex number. A piecewise constant persistently exciting (PCPE) input of order $L$ for system \eqref{ctsys} is defined as ${u(t + iT) = \mu_i}$ for all $0 \leq t < T$, $i=0,\ldots,N-1$, where the sequence of constant vectors $\{ \mu_i \}_{i=0}^{N-1}$, $\mu_i \in \mathbb{R}^m$, satisfies
	\begin{equation*}
		\text{rank} \left( \left[ \begin{array}{cccc}
			\mu_0 & \mu_1 & \cdots & \mu_{N-L} \\
			\vdots & \vdots & \ddots & \vdots \\
			\mu_{L-1} & \mu_{L} & \cdots & \mu_{N-1}
		\end{array} \right] \right) = mL.
	\end{equation*}
\end{defn}

Condition (\ref{assut}) imposes restrictions on the time interval $T$. Although (\ref{assut}) uses model knowledge, it can be observed that the values of $T$ that do not satisfy (\ref{assut}) form a set of measure zero and, therefore, an arbitrary selection of $T$ is almost surely useful for Definition \ref{defctpe}.

\begin{rem}
	Throughout this paper we assume that a piecewise constant input as in Definition~\ref{defctpe} is used to excite the system\footnote{ A PCPE input as in Definition~\ref{defctpe} is convenient as it is known to satisfy the rank condition \eqref{pecond}. Other PE inputs can also be used to excite system \eqref{ctsys}, as long as \eqref{pecond} holds.}. Then, the matrix $\mathcal{H}_{\mathcal{T}}(u(t))$ as in (\ref{hankrow}) is constant in $t$ and we denote it as $\mathcal{H}_{\mathcal{T}}(u) := \mathcal{H}_{\mathcal{T}}(u(t))$.
\end{rem}

The following lemma, which is a special case of \cite[Lemma 1]{LopezMuCDC2022}, shows that the data collected from system (\ref{ctsys}) after the application of a PCPE input satisfies an important excitation condition.

\begin{lem}
	\label{lempe}
	Consider system (\ref{ctsys}), let the pair $(A,B)$ be controllable, and let $u$ be a PCPE input of order $n+1$ as defined in Definition \ref{defctpe}. Then, for all $0 \leq t \leq T$ it holds that
	\begin{equation}
		\text{rank} \left( \left[ \begin{array}{c}
			\mathcal{H}_{\mathcal{T}}(u) \\ \mathcal{H}_{\mathcal{T}}(x_{f}(t))
		\end{array} \right] \right) = m + n.
		\label{pecond}
	\end{equation}
\end{lem}

In \cite{LopezMuCDC2022}, a continuous-time version of Willems' fundamental lemma \citep{WillemsRapMarDe2005} was introduced. Lemma~\ref{lemdifeq} and Theorem~\ref{thmctwill} below, which are special cases of \cite[Lemma~2]{LopezMuCDC2022} and \cite[Theorem~2]{LopezMuCDC2022}, describe this result for the particular case when PCPE inputs are used to excite the system and state measurements are available. This is the setting that we consider in the remainder of this paper.

\begin{lem}
	\label{lemdifeq}
	Consider a controllable system (\ref{ctsys}). Let $u : [0,NT] \rightarrow \mathbb{R}^m$, $T > 0$, $N \in \mathbb{N}$, be a PCPE input of order $n+1$, and let $x_{f} : [0,NT] \rightarrow \mathbb{R}^n$, be the corresponding state of (\ref{ctsys}). Moreover, consider an arbitrary continuously differentiable signal $\bar u : [0,T] \rightarrow \mathbb{R}^m$ and let $\bar x(0) \in \mathbb{R}^n$ be an arbitrary vector. Then, there exists a solution $\alpha(t)$ for $0 \leq t \leq T$ of the differential equation
	\begin{equation}
		\left[ \begin{array}{c}
			\mathcal{H}_{\mathcal{T}}(u) \\ \mathcal{H}_{\mathcal{T}}(x_{f}(t))
		\end{array} \right] \dot \alpha(t) = \left[ \begin{array}{c}
			\dot{\bar{u}}(t) \\ 0
		\end{array} \right],
		\label{difeqa}
	\end{equation}
	with initial condition constraint
	\begin{equation}
		\left[ \begin{array}{c}
			\mathcal{H}_{\mathcal{T}}(u) \\ \mathcal{H}_{\mathcal{T}}(x_{f}(0))
		\end{array} \right] \alpha(0) = \left[ \begin{array}{c}
			\bar{u}(0) \\ \bar x(0)
		\end{array} \right].
		\label{inicoa}
	\end{equation}
\end{lem}

\begin{thm}
	\label{thmctwill}
	Let the conditions in Lemma \ref{lemdifeq} hold. Then, any signals $\bar u : [0,T] \rightarrow \mathbb{R}^m$, $\bar x : [0,T] \rightarrow \mathbb{R}^n$, where $\bar u$ is continuously differentiable, are an input-state trajectory of (\ref{ctsys}) corresponding to some initial condition $\bar x(0)$ if and only if there exists a continuously differentiable vector function $\alpha$ such that the equations (\ref{difeqa}), (\ref{inicoa}) and
	\begin{equation}
		\left[ \begin{array}{c}
			\mathcal{H}_{\mathcal{T}}(u) \\ \mathcal{H}_{\mathcal{T}}(x_{f}(t))
		\end{array} \right] \alpha(t) = \left[ \begin{array}{c}
			\bar u(t) \\ \bar x(t)
		\end{array} \right]
		\label{ctwillem2}
	\end{equation}
	hold for $0 \leq t \leq T$.
\end{thm}

Theorem \ref{thmctwill} states that any input-state trajectory $\bar{u}$, $\bar{x}$ of (\ref{ctsys}) can be represented via (\ref{ctwillem2}) using one persistently excited data trajectory $u$, $x$ collected from \eqref{ctsys}. In the following section, this result is used to analyze the conditions for a gain matrix $K$ to stabilize the system.

\section{Conditions for stabilizing control of continuous-time systems}
\label{seccond}

In this section we provide a framework for stability analysis of a closed-loop system using the data-based system representation \eqref{difeqa}-\eqref{ctwillem2}. These developments will be exploited in Section \ref{sectr} for control design. Although the results in the previous section were obtained for noise-free data, it is of practical interest to consider here noisy measurements of the state and its derivative as
\begin{equation}
	x(t) := x_{f}(t) + \varepsilon_1(t), \quad \dot x(t) := \dot x_{f}(t) + \varepsilon_2(t)
	\label{noisymes}
\end{equation}
 for some noise terms $\varepsilon_1$, $\varepsilon_2$. If $\dot x$ is estimated from $x$, its expression in \eqref{noisymes} accounts for approximation errors. In the following, we make the following assumption.

\begin{assum}
	\label{assnpe}
	Let $u : [0,NT] \rightarrow \mathbb{R}^m$, $T > 0$, $N \in \mathbb{N}$, be a PCPE input of order $n+1$, and let $x : [0,NT] \rightarrow \mathbb{R}^n$, $\dot x : [0,NT] \rightarrow \mathbb{R}^n$ be the noisy state and state derivative measurements of the controllable system \eqref{ctsys}, as in \eqref{noisymes}. Then, $(u,x)$ are such that \eqref{pecond} holds with $x_f$ replaced by $x$.
\end{assum}

Note that, by Lemma \ref{lempe}, Assumption \ref{assnpe} holds with probability $1$ in the case of white noise, or when the magnitude of the noise is small enough. Moreover, Assumption~\ref{assnpe} is easy to verify from data.

The following result provides conditions for stability of a closed-loop system. The goal of this theorem is not control design, which is studied in the next section. Instead, the objective is to state necessary and sufficient conditions for stability that recover other conditions in the literature, as shown afterwards.

\begin{thm}
	\label{thmstab}
	Consider system \eqref{ctsys} and let Assumption~\ref{assnpe} hold. Consider an arbitrary $K \in \mathbb{R}^{m \times n}$ and an arbitrary $t \in [0,T]$. There exists a $\Gamma \in \mathbb{R}^{N \times n}$ such that
	\begin{IEEEeqnarray}{c}
		\Bigl( \mathcal{H}_{\mathcal{T}}(u) + K \mathcal{H}_{\mathcal{T}}(x(t)) \Bigl) \Gamma = 0, \label{constabu}\\
		\mathcal{H}_{\mathcal{T}}(x(t)) \Gamma \text{ is nonsingular}. \label{constabx}
	\end{IEEEeqnarray}
	 The closed-loop matrix $A-BK$ is Hurwitz if and only if
	\begin{equation}
		\Bigl( \mathcal{H}_{\mathcal{T}}(\dot x(t)) + \mathcal{H}_{\mathcal{T}}(v(t)) \Bigl) \Gamma \Bigl( \mathcal{H}_{\mathcal{T}}(x(t)) \Gamma \Bigl)^{-1} \text{ is Hurwitz}, \label{constabdx}
	\end{equation}
	where $\Gamma$ satisfies \eqref{constabu}-\eqref{constabx}, and $v(t):=A \varepsilon_1(t) - \varepsilon_2(t)$.
\end{thm}

\begin{pf}
As in the theorem statement, consider any $t \in [0,T]$. Throughout this proof this time instant remains fixed. To show the existence of the matrix $\Gamma$, let $\bar x_{1}$ be an arbitrary state of \eqref{ctsys} at time $t$. By Assumption \ref{assnpe}, $\mathcal{H}_{\mathcal{T}}(x(t))$ has full row rank and there exists a vector $\alpha_1 \in \mathbb{R}^N$ such that $\mathcal{H}_{\mathcal{T}}(x(t)) \alpha_1 = \bar x_{1}$. Using the same argument for $n$ linearly independent vectors $\bar x_{i}$, $i=1,\ldots,n$, we can always determine $\alpha_i \in \mathbb{R}^N$ such that
\begin{equation}
	\mathcal{H}_{\mathcal{T}}(x(t)) \Gamma = X,
	\label{xgamma}
\end{equation}
for the fixed $t$, where $\Gamma := \left[ \alpha_1 \quad \alpha_2 \quad \cdots \quad \alpha_n \right]$ and $X := \left[ \bar x_{1} \quad \bar x_{2} \quad \cdots \quad \bar x_{n} \right]$. Moreover, by Assumption \ref{assnpe} the matrix $\left[ \begin{array}{c} \mathcal{H}_{\mathcal{T}}(u) + K \mathcal{H}_{\mathcal{T}}(x(t)) \\ \mathcal{H}_{\mathcal{T}}(x(t)) \end{array} \right]$ has full row rank for any $K \in \mathbb{R}^{m \times n}$ and any $t \in [0,T]$. Thus, $\Gamma$ can be chosen such that, in addition to \eqref{xgamma}, also \eqref{constabu} is satisfied. Since $X$ is nonsingular, the matrix $\Gamma$ is now as in \eqref{constabu}-\eqref{constabx}.

We now show that \eqref{constabdx} is a sufficient and necessary condition for the closed-loop matrix $A-BK$ to be Hurwitz. Consider the expression \eqref{hdyn} and multiply on the right by $\Gamma$ to obtain $\mathcal{H}_{\mathcal{T}}(\dot x_f(t)) \Gamma = A \mathcal{H}_{\mathcal{T}}(x_f(t))\Gamma + B \mathcal{H}_{\mathcal{T}}(u) \Gamma$. Solving for $x_{f}$ and $\dot x_{f}$ from \eqref{noisymes} and substituting in this expression, we can write $\mathcal{H}_{\mathcal{T}}(\dot x(t))\Gamma + \mathcal{H}_{\mathcal{T}}(v(t)) \Gamma = A \mathcal{H}_{\mathcal{T}}(x(t)) \Gamma + B \mathcal{H}_{\mathcal{T}}(u) \Gamma$, where $v$ is as in the theorem statement. Using \eqref{constabu} here, we get
\begin{equation}
	\Bigl( \mathcal{H}_{\mathcal{T}}(\dot x(t)) + \mathcal{H}_{\mathcal{T}}(v(t)) \Bigl) \Gamma = (A-BK) \mathcal{H}_{\mathcal{T}}(x(t)) \Gamma.
	\label{clgamma}
\end{equation}
Condition (\ref{constabx}) implies that, from \eqref{clgamma}, we can write
\begin{equation}
	A-BK = \Bigl( \mathcal{H}_{\mathcal{T}}(\dot x(t)) + \mathcal{H}_{\mathcal{T}}(v(t)) \Bigl) \Gamma \left( \mathcal{H}_{\mathcal{T}}(x(t)) \Gamma \right)^{-1}
	\label{clabk}
\end{equation}
and $K$ is stabilizing if and only if (\ref{constabdx}) holds. $\square$
\end{pf}

Since the signal $v$ is unknown, condition \eqref{constabdx} cannot directly be used for control design. However, Theorem~\ref{thmstab} provides conditions related to the measurement noise that guarantee stability of the system. In the following, we show that \eqref{constabu}-\eqref{constabdx} are general expressions that can be used to recover other results in the literature\footnote{Different from some results in the literature, in this paper we use the negative feedback standard notation $u=-Kx$. This results in sign differences in our notation.}.

In \cite{EisingCor2023}, the authors report conditions for informativity and stability of disturbed CT systems of the form
\begin{equation}
	\dot x(t) = Ax(t) + Bu(t) + w(t),
	\label{nsys}
\end{equation}
where, given $\bar W \succ 0$, the disturbances $w \in \mathbb{R}^n$ satisfy
\begin{equation}
	T \mathcal{H}_{\mathcal{T}}(w(t)) \mathcal{H}_{\mathcal{T}}(w(t))^\top \preceq \bar W, \quad 0 \leq t \leq T.
	\label{noisecond}
\end{equation}
Namely, $u=-Kx$ stabilizes (\ref{nsys}) if and only if there is a scalar $\beta > 0$ and matrices $P \in \mathbb{R}^{n \times n}$, $L \in \mathbb{R}^{m \times n}$ such that, for a fixed $t \in [0,T]$ and $\bar W_\beta := \bar W + \beta I$,
\begin{IEEEeqnarray}{c}
	T \left[ \begin{array}{c}
		\mathcal{H}_{\mathcal{T}}(\dot x(t)) \\ -\mathcal{H}_{\mathcal{T}}(x(t)) \\ -\mathcal{H}_{\mathcal{T}}(u)
	\end{array} \right] \left[ \begin{array}{c}
		\mathcal{H}_{\mathcal{T}}(\dot x(t)) \\ -\mathcal{H}_{\mathcal{T}}(x(t)) \\ -\mathcal{H}_{\mathcal{T}}(u)
	\end{array} \right]^\top 
	- \left[ \begin{array}{ccc}
		\bar W_\beta & P & L^\top \\ P & 0 & 0 \\ L & 0 & 0
	\end{array} \right] \succeq 0, \nonumber \\
	P \succ 0, \quad \text{and} \quad K=-LP^{-1}.
	\label{cortes}
\end{IEEEeqnarray}
		 
We show that, starting from (\ref{constabu})-(\ref{constabdx}), the stability conditions in \eqref{cortes} can be obtained. First, notice from \eqref{ctsys} and \eqref{noisymes} that the noisy data satisfy $\dot x(t) = Ax(t) + Bu(t) - v(t)$,	with $v$ as in Theorem \ref{thmstab}. Comparing with \eqref{nsys}, let \eqref{noisecond} hold with $w$ replaced by $v$. Now, let $\Gamma$ satisfy
\begin{equation}
	\left[ \begin{array}{c} \mathcal{H}_{\mathcal{T}}(u) \\ \mathcal{H}_{\mathcal{T}}(x(t)) \end{array} \right] \Gamma = \left[ \begin{array}{c} -KP \\ P \end{array} \right] =: \left[ \begin{array}{c} L \\ P \end{array} \right]
	\label{stabec2}
\end{equation}
for some $P \succ 0$, such that \eqref{constabu}-\eqref{constabx} hold. This implies also the last two conditions in \eqref{cortes}. Finally, from \eqref{constabdx} and \eqref{clabk} it follows that $P$ in \eqref{stabec2} can be chosen such that $(A-BK)P + P(A-BK)^\top \prec 0$. From this point, the same procedure as in \cite{EisingCor2023} can be followed to obtain the first inequality in \eqref{cortes}.

Many data-based conditions for stability of linear systems in the literature have been proposed for noise-free data, which can be compared to the conditions \eqref{constabu}-\eqref{constabdx} with $v\equiv 0$. For example, \cite[Remark~2]{DePersisTes2020} presented the first data-based conditions for stability of CT systems as follows. Using our notation, $K$ is stabilizing if, for a fixed $t \in [0,T]$, it can be written as
\begin{equation}
	K = - \mathcal{H}_{\mathcal{T}}(u) \Gamma (\mathcal{H}_{\mathcal{T}}(x(t)) \Gamma )^{-1}, \label{stabdp1}
\end{equation}
where the matrix $\Gamma \in \mathbb{R}^{N \times n}$ satisfies the LMIs 
\begin{IEEEeqnarray}{c}
	\mathcal{H}_{\mathcal{T}}(x(t)) \Gamma \succ 0, \label{stabdp2} \\
	\mathcal{H}_{\mathcal{T}}(\dot x(t)) \Gamma + \Gamma^\top \mathcal{H}_{\mathcal{T}}(\dot x(t))^\top \prec 0. \label{stabdp3}
\end{IEEEeqnarray}
Note that (\ref{constabu}) together with (\ref{constabx}) allows to express $K$ as in (\ref{stabdp1}). Also, (\ref{stabdp2}) is a particular case of (\ref{constabx}). To see that \eqref{stabdp3} is a special case of \eqref{constabdx} for $v \equiv 0$, let $\mathcal{H}_{\mathcal{T}}(x(t)) \Gamma = P$ for some $P \succ 0$, and write (\ref{stabdp3}) as
\begin{multline}
	\mathcal{H}_{\mathcal{T}}(\dot x(t)) \Gamma \Bigl( \mathcal{H}_{\mathcal{T}}(x(t)) \Gamma \Bigr)^{-1}P \\
	+ P \Bigl( \Gamma^\top \mathcal{H}_{\mathcal{T}}(x(t))^\top \Bigr)^{-1} \Gamma^\top \mathcal{H}_{\mathcal{T}}(\dot x(t))^\top \prec 0.
	\label{constablyap}
\end{multline}

Other stability conditions can be obtained by choosing $\Gamma$ differently. For example, consider the conditions
\begin{IEEEeqnarray}{c}
	K = - \mathcal{H}_{\mathcal{T}}(u) \Gamma, \\
	\mathcal{H}_{\mathcal{T}}(x(t)) \Gamma = I, \label{stabvw2}\\
	\mathcal{H}_{\mathcal{T}}(\dot x(t)) \Gamma \text{ is Hurwitz},
\end{IEEEeqnarray}
for some fixed $t \in [0,T]$, which are also special cases of (\ref{constabu})-(\ref{constabdx}) for $v \equiv 0$. Notice that these expressions are analogous to the conditions for stability of discrete-time systems obtained in \cite[Theorem 16]{vanWaardeEisTreCam2020}, where $\Gamma$ takes the role of a right inverse of $\mathcal{H}_{\mathcal{T}}(x(t))$.

\begin{rem}
	Different from the conditions in the literature, the developments in the proof of Theorem~\ref{thmstab} provide a novel \emph{meaning} to the matrix $\Gamma$ (compare \eqref{xgamma}). In our approach, $\Gamma$ is seen as a trajectory-generating matrix for system (\ref{ctsys}) as described by (\ref{ctwillem2}). This establishes a link between the stability conditions (\ref{constabu})-(\ref{constabdx}) and the data-based system representation in Lemma \ref{lemdifeq} and Theorem \ref{thmctwill}.
\end{rem} 

In the following sections, we exploit the conditions (\ref{constabu})-(\ref{constabdx}), as well as the expressions (\ref{clgamma}) and (\ref{clabk}), to design controllers with different performance requirements.

\section{Data-based control design with performance specifications}
\label{sectr}

In this section we present our solutions for three continuous-time data-based control problems: trajectory-reference control, linear quadratic regulation, and robust pole placement.

\subsection{Trajectory-reference control}

The objective in the trajectory-reference problem is to use an available set of desired state trajectories to design a control policy that allows the system (\ref{ctsys}) to follow them as closely as possible. This problem is encountered when specific transient performances are desired, or when trajectories from an expert system to be emulated are available. In this paper, we focus on stabilizing the system to the origin while following desired transient performances. This is formalized as follows.

\begin{prob}
	\label{probbc}
	Let $\xi^i: [0,T] \rightarrow \mathbb{R}^n$, $i=1,\ldots,M$, be a set of $M \geq 1$ trajectories, and let their known time derivatives be $\dot \xi^i$, $i=1,\ldots,M$. Define the matrices
	\begin{equation}
		\Xi(t) = \left[ \xi^1(t) \quad \xi^2(t) \quad \cdots \quad \xi^M(t) \right],
		\label{exptraj}
	\end{equation}
	\begin{equation}
		\dot \Xi(t) = \left[ \dot \xi^1(t) \quad \dot \xi^2(t) \quad \cdots \quad \dot \xi^M(t) \right],
		\label{expder}
	\end{equation}
	for $0 \leq t \leq T$. Using data as in Assumption \ref{assnpe}, determine a state feedback matrix $K$ such that $A-BK$ is Hurwitz and, for each $i = 1,\ldots, M$, when $x(0) = \xi^i(0)$ the error integral $\int_0^T \| x_f(\tau) - \xi^i(\tau) \|_F \, d\tau$ is minimized.
\end{prob}

To solve Problem~\ref{probbc}, we exploit persistently excited data such that the conditions in Lemma~\ref{lemdifeq} and Theorem~\ref{thmctwill} hold. Hence notice that, from \eqref{ctwillem2} (compare also \eqref{xgamma}), for every set of $M$ state trajectories, there exists a matrix $\Gamma(t)$, $0 \leq t \leq T$, such that
\begin{equation}
	\left[ \bar x_f^1(t) \quad \cdots \quad \bar x_f^M(t) \right] =: X(t) = \mathcal{H}_{\mathcal{T}}(x_f(t)) \Gamma(t).
	\label{xnsquare}
\end{equation}
For $\Gamma(t)$ to generate state feedback trajectories, it must also satisfy\footnote{Note that the presence of noise may not allow for these equations to hold exactly. This is taken into account in our final solution to Problem~\ref{probbc} below.} $\mathcal{H}_{\mathcal{T}}(u) \Gamma(t) = -K X(t)$, as well as the differential equation (\ref{difeqa}), that is $\mathcal{H}_{\mathcal{T}}(u) \dot \Gamma(t) = -K \dot X(t)$ and $\mathcal{H}_{\mathcal{T}}(x_f(t)) \dot \Gamma(t) = 0$. Finally, it is desired that $X(t) \approx \Xi(t)$ for $0 \leq t \leq T$. Notice that, different from (\ref{xgamma}), $X(t) \in \mathbb{R}^{n \times M}$ in \eqref{xnsquare} is not necessarily a square matrix. 

From these expressions, a \emph{naive} optimization problem could be formulated as follows to solve Problem \ref{probbc}
\begin{subequations}
	\begin{align}
		& \underset{\Gamma(t),\bar K}{\text{minimize}} \quad \int_0^T \| \mathcal{H}_{\mathcal{T}}(x_f(\tau)) \Gamma(\tau) - \Xi(\tau) \|_F \, d\tau \label{naive1}\\ 
		& \text{s.t.} \qquad \mathcal{H}_{\mathcal{T}}(x_f(0)) \Gamma(0) = \Xi(0), \label{naive2}\\
		& \qquad \quad \, \mathcal{H}_{\mathcal{T}}(u) \Gamma(0) = -\bar K \Xi(0), \label{naive22}\\
		& \left[ \begin{array}{c} \mathcal{H}_{\mathcal{T}}(u) \\ \mathcal{H}_{\mathcal{T}}(x_f(t)) \end{array} \right] \dot \Gamma(t) = \left[ \begin{array}{c} -\bar K \mathcal{H}_{\mathcal{T}}(\dot x_f(t)) \Gamma(t) \\ 0 \end{array} \right], \label{naive3} \\
		& \quad \text{for } 0 \leq t \leq T. \nonumber
	\end{align}
\label{optprobnaiv}
\end{subequations}
The constraints (\ref{naive2}) and (\ref{naive22}) set the initial conditions of the desired trajectories and the state feedback input, respectively. The cost (\ref{naive1}) minimizes the error between the desired trajectories $\Xi(t)$ and the system trajectories that evolve in time according to the differential equation (\ref{naive3}). The problem (\ref{optprobnaiv}) is modified as follows to construct a convex program applicable in practice. 

First, we can avoid the need for the explicit use of (\ref{naive3}) by replacing it with the equations $\mathcal{H}_{\mathcal{T}}(x_f(t)) \dot \Gamma(t) = 0$ and
\begin{equation}
	\mathcal{H}_{\mathcal{T}}(u) \Gamma(t) = -\bar K \mathcal{H}_{\mathcal{T}}(x_f(t)) \Gamma(t)
	\label{algequ}
\end{equation}
for $0 \leq t \leq T$. This is because, given $\mathcal{H}_{\mathcal{T}}(x_f(t)) \dot \Gamma(t) = 0$, as well as the initial conditions set by \eqref{naive2}-\eqref{naive22}, the equation $\mathcal{H}_{\mathcal{T}}(u) \dot \Gamma(t) = -\bar K \mathcal{H}_{\mathcal{T}}(\dot x_f(t)) \Gamma(t)$ is satisfied for $0 \leq t \leq T$ if and only if \eqref{algequ} holds. To see this, the time derivative of (\ref{algequ}) can be taken. 
Notice also that the desired objective $\mathcal{H}_{\mathcal{T}}(x_f(t)) \Gamma(t) = \Xi(t)$ implies $\frac{d}{dt} \left( \mathcal{H}_{\mathcal{T}}(x_f(t)) \Gamma(t) \right) = \dot \Xi(t)$. Thus, if the equations
\begin{IEEEeqnarray}{c}
	\mathcal{H}_{\mathcal{T}}(x_f(t)) \Gamma(t) = \Xi(t), \label{algeqx} \\
	\mathcal{H}_{\mathcal{T}}(\dot x_f(t)) \Gamma(t) = \dot \Xi(t) \label{algeqdx}
\end{IEEEeqnarray}
hold for $0 \leq t \leq T$, then $\mathcal{H}_{\mathcal{T}}(x_f(t)) \dot \Gamma(t) = 0$ as desired. Finally, although continuous signals are considered in Problem~\ref{probbc}, in practice only samples of such trajectories can be stored and manipulated in digital computers. If there are samples available at times $\{ t_1, \, t_2,\ldots, t_{q}  \}$, where $0 \leq t_i \leq T$ for $i=1,\ldots,q$, then using \eqref{algequ}-\eqref{algeqdx} we obtain the following approximation of problem (\ref{optprobnaiv})
\begin{subequations}
	\begin{align}
			\underset{\Gamma(t_i),\bar K}{\text{minimize}} \quad & \sum_{i = 1}^{q}  \| \mathcal{H}_{\mathcal{T}}(\dot x(t_i)) \Gamma(t_i) - \dot \Xi(t_i) \|_F \label{costbc} \\
			\text{s.t.} \quad & \mathcal{H}_{\mathcal{T}}(u) \Gamma(t_i) = - \bar K \Xi(t_i), \label{constbc1}\\
			& \mathcal{H}_{\mathcal{T}}(x(t_i)) \Gamma(t_i) = \Xi(t_i), \; i=1,\ldots,q. \label{constbc2}
		\end{align}%
\label{optprobbc}
\end{subequations}%

Instead of an integral as in \eqref{naive1}, a sum over the known trajectory samples is used. The constraints (\ref{constbc1})-(\ref{constbc2}) imply that \eqref{algequ}-\eqref{algeqx} hold for all samples. Condition \eqref{algeqdx} is used as a soft constraint (i.e., appearing in the cost \eqref{costbc}) since, in general, $\mathcal{H}_{\mathcal{T}}(\dot x(t_i)) \Gamma(t_i) \neq \dot \Xi(t_i)$. This is because arbitrary desired trajectories $\Xi$ might not be tractable. In contrast, \eqref{constbc1} and \eqref{constbc2} are feasible for all $t_i$ by Assumption~\ref{assnpe}. The following theorem states conditions for the optimal $\bar K$ to allow exact tracking of the trajectories in $\Xi(t)$. Note that one of these conditions is the availability of noise-free data.

\begin{thm}
	\label{thmtr}
	Consider the controllable system (\ref{ctsys}). Let $u$ be a PCPE input of order $n+1$, and let $x$, $\dot x$, be the corresponding state and state derivative measurements \eqref{noisymes} with $\varepsilon_1 \equiv \varepsilon_2 \equiv 0$. Moreover, consider the desired trajectories in \eqref{exptraj}-\eqref{expder}. If a matrix $K$ exists such that
	\begin{equation}
		\dot \Xi(t) = (A-B K) \Xi(t), \quad 0\leq t \leq T,
		\label{zerocostk}
	\end{equation}
	then the solution $\bar K$ of \eqref{optprobbc} satisfies \eqref{zerocostk} and the optimal cost \eqref{costbc} equals zero.
\end{thm}
\begin{pf}
	If a matrix $K$ as in \eqref{zerocostk} exists, then $\Xi(t)$ can be generated by the data-based system representation in Theorem \ref{thmctwill}. That is, there exists a $\Gamma(t)$ such that \eqref{algequ}-\eqref{algeqdx} hold for $0 \leq t \leq T$. Thus, selecting $\bar K$ in \eqref{optprobbc} as in \eqref{zerocostk} makes the cost \eqref{costbc} equal to zero. Now, we show that if the cost of \eqref{optprobbc} is zero, then $\bar K$ satisfies \eqref{zerocostk}. This is obtained using similar arguments as in the proof of Theorem~\ref{thmstab} by noticing that a zero cost implies $\mathcal{H}_{\mathcal{T}}(\dot x(t_i)) \Gamma(t_i) = (A-B \bar K) \mathcal{H}_{\mathcal{T}}(x(t_i)) \Gamma(t_i)$ (compare to (\ref{clgamma})). Substituting $\mathcal{H}_{\mathcal{T}}(x(t_i)) \Gamma(t_i) = \Xi(t_i)$ and $\mathcal{H}_{\mathcal{T}}(\dot x(t_i)) \Gamma(t_i) = \dot \Xi(t_i)$ in this expression yields $\dot \Xi(t) = (A-B \bar K) \Xi(t)$, completing the proof. $\square$
\end{pf}

\begin{rem}
	The need in Theorem \ref{thmtr} for noise-free data is only for showing that a matrix $K$ as in \eqref{zerocostk} can be obtained. Problem \eqref{optprobbc} is feasible and can be solved with noisy measurements under Assumption 6.
\end{rem}

Constraining $\bar K$ in (\ref{optprobbc}) to be stabilizing is a challenging task. Instead, we can obtain a stabilizing $\bar K$ in a separate step after solving (\ref{optprobbc}). Given a matrix $\bar K$, we aim to determine a stabilizing matrix $K$ that renders the closed-loop matrix $A-BK$ as close as possible to $A-B\bar K$. If $\bar K$ was already stabilizing, we do not wish to modify it. This objective can be achieved as follows. Given the matrix $\bar K$ and a fixed time instant $t \in [0,T]$, solve
\begin{subequations}
	\begin{align}
		\underset{G_1,G_2,P,L,\beta}{\text{minimize}} \qquad & \| \mathcal{H}_{\mathcal{T}}(\dot x(t)) (G_1 - G_2) \|_F \label{costbc2}\\ 
		\text{s.t.} \qquad &  (\ref{cortes}) \text{ holds}, \\
		& \mathcal{H}_{\mathcal{T}}(x(t)) G_1 = P,  \label{constbc21} \\
		& \mathcal{H}_{\mathcal{T}}(u) G_1 = L, \label{constbc22} \\
		& \mathcal{H}_{\mathcal{T}}(x(t)) G_2 = P, \label{constbc23} \\
		& \mathcal{H}_{\mathcal{T}}(u) G_2 = - \bar K P. \label{constbc24}
	\end{align}
	\label{optprobbc2}
\end{subequations}
Then, let $K = -L P^{-1}$. In \eqref{optprobbc2}, we make use of the stabilizing conditions \eqref{cortes} which allow to handle noisy data (see Section~\ref{seccond}). The following theorem shows the properties of the solution $K$ obtained from (\ref{optprobbc2}).

\begin{thm}
	\label{thmtr2}
	Consider system \eqref{ctsys}, let Assumption~\ref{assnpe} hold, and let \eqref{noisecond} hold with $w$ replaced by $v:=A \varepsilon_1 - \varepsilon_2$. Consider also a matrix $\bar K$ and a fixed time $t \in [0,T]$. The solution $K$ of (\ref{optprobbc2}) is stabilizing and such that the difference $\| (A-BK)P - (A-B\bar K)P -\mathcal{H}_{\mathcal{T}}(v(t)) (G_1-G_2) \|_F$ is minimized. Moreover, if $\bar K$ is stabilizing, then $K=\bar K$.
\end{thm}
\begin{pf}
	The fact that $K$ is stabilizing follows as in \cite{EisingCor2023}. The constraints (\ref{constbc21}) and (\ref{constbc22}) correspond to the condition (\ref{stabec2}). Notice that, since $L=-KP$, we can write $\mathcal{H}_{\mathcal{T}}(u) G_1 = L = -KP = -K\mathcal{H}_{\mathcal{T}}(x(t)) G_1$. From the discussion in Section \ref{seccond} (compare (\ref{constabu}) and the proof of Theorem \ref{thmstab}), this implies that (\ref{clgamma}) holds with $\Gamma$ replaced by $G_1$. By (\ref{constbc21}), (\ref{clgamma}) can be rewritten as $\left( \mathcal{H}_{\mathcal{T}}(\dot x(t)) + \mathcal{H}_{\mathcal{T}}(v(t)) \right) G_1 = (A-BK)P$. The constraints (\ref{constbc23}) and (\ref{constbc24}) have the same structure as (\ref{constbc21}) and (\ref{constbc22}), except that the use of $G_2$ replaces $K$ by $\bar K$. Therefore, we obtain $\left( \mathcal{H}_{\mathcal{T}}(\dot x(t)) + \mathcal{H}_{\mathcal{T}}(v(t)) \right) G_2 = (A-B\bar K)P$. This shows that the cost (\ref{costbc2}) is equivalent to $\| (A-BK)P - (A-B\bar K)P -\mathcal{H}_{\mathcal{T}}(v(t)) (G_1-G_2) \|_F$ as claimed. Finally, if $\bar K$ is stabilizing, then it can be written as $\bar K = -L P^{-1}$ for some $L$ and $P$ that satisfy the conditions \eqref{cortes} \citep{EisingCor2023}. Thus, the optimal solution to (\ref{optprobbc2}) is such that $G_1 = G_2$, leading to $K = \bar K$. $\square$
\end{pf}

Thus, our solution to Problem \ref{probbc} consists of the following steps: (i) Determine $\bar K$ by solving (\ref{optprobbc}). (ii) Using $\bar K$, solve the optimization problem (\ref{optprobbc2}). (iii) Using $P$ and $L$, compute the solution $K = L P^{-1}$.

\subsection{Data-based optimal control for CT systems}
\label{secoc}

In this section we solve the data-based LQR problem for CT systems. These results are obtained for the case of noise-free data, i.e., $\varepsilon_1 \equiv \varepsilon_2 \equiv 0$ in \eqref{noisymes}. In Section~\ref{secsim}, we show in simulation examples that our solution has inherent robustness properties for small magnitudes of noise. Thus, the problem description is as follows.

 \begin{prob}
	For every initial condition of system (\ref{ctsys}), use noise-free measured data to determine the control input $u$ that minimizes the cost
	\begin{equation}
		\int_0^\infty \Bigl( x(t)^\top Q x(t) + u(t)^\top R u(t) \Bigl) dt
		\label{cost}
	\end{equation}
	with $Q \succeq 0$ such that $(A,Q^{1/2})$ is observable, and $R \succ 0$.
\end{prob}

For discrete-time systems, several solutions to the data-based LQR problem have been proposed (see, e.g., \cite{DePersisTes2020,vanWaardeEisTreCam2020}). To the best of our knowledge, for continuous-time systems the only data-based solutions in the literature are reinforcement learning algorithms \citep{JiangJia2012,LopezMulCDC2023}. In \cite{LopezMulCDC2023} it is shown that these methods have attractive features (e.g., their computational complexity), but numerical issues with the required solvers for high-dimensional systems were reported. Here, we show that an algorithm analogous to the one in \cite{vanWaardeEisTreCam2020} for DT systems can also be developed for the CT case, providing a method that does not require to solve the matrix equations in \cite{LopezMulCDC2023}.
 
The LQR solution is given by $K^* = R^{-1} B^\top P^*$, where $P^* \succ 0$ is the unique positive definite solution of the algebraic Riccati equation (ARE) \cite[Theorem~4.1]{Wonham1968}
 \begin{equation}
 	Q + P^*A + A^\top P^* - P^*BR^{-1} B^\top P^* = 0.
 	\label{riccati}
 \end{equation}
 Thus, it follows that the matrices $K^*$ and $P^*$ satisfy
 \begin{equation}
 	Q + K^{*\top} R K^* + P^*(A-BK^*) + (A-BK^*)^\top P^* = 0.
 	\label{bellmaneq}
 \end{equation}
 The following result is straightforwardly obtained.
 
 \begin{lem}
 	\label{lempmax}
 	Consider system (\ref{ctsys}) and the cost function (\ref{cost}). The solution $P^* \succ 0$ of (\ref{riccati}) is such that $P^* \succeq P$ for any matrix $P \succ 0$ that satisfies 
 	 \begin{equation}
 		Q + K^{*\top} R K^* + P(A-BK^*) + (A-BK^*)^\top P \succeq 0.
 		\label{bellmanineq}
 	\end{equation}
 \end{lem}
 \begin{pf}
 	From (\ref{bellmaneq}) and (\ref{bellmanineq}) it follows that
 	\begin{multline*}
 		P(A-BK^*) + (A-BK^*)^\top P \succeq \\
 		P^*(A-BK^*) + (A-BK^*)^\top P^*,
 	\end{multline*}
 	implying $(P^* - P)(A-BK^*) + (A-BK^*)^\top (P^* - P) \preceq 0$. Now, since $A-BK^*$ is Hurwitz, an equation of the form $\mathcal{P} (A-BK^*) + (A-BK^*)^\top \mathcal{P} = - \mathcal{Q}$ has a unique solution $\mathcal{P}$ given by $\mathcal{P} = \int_0^\infty e^{(A-BK^*)^\top t} \mathcal{Q} e^{(A-BK^*)t} dt$ \cite[Theorem~5.5]{Chen1999}. Clearly, if $\mathcal{Q} \succeq 0$, then also $\mathcal{P} \succeq 0$, which in this case implies $P^* - P \succeq 0$. $\square$
 \end{pf}

 To solve the LQR problem for CT systems, solve first the following optimization problem for a fixed $t \in [0,T]$
 \begin{align}
 	\underset{P}{\text{maximize}} \quad & \trace (P) \label{optproblqr} \\
 	\text{s.t.} \qquad & P\succ 0, \quad \mathcal{L}(P) \succeq 0, \nonumber 
 \end{align}
 where
 \begin{multline}
 	\mathcal{L}(P) = \mathcal{H}_{\mathcal{T}}(x(t))^\top Q \mathcal{H}_{\mathcal{T}}(x(t)) + \mathcal{H}_{\mathcal{T}}(u)^\top R \mathcal{H}_{\mathcal{T}}(u) \\
 	+ \mathcal{H}_{\mathcal{T}}(x(t))^\top P \mathcal{H}_{\mathcal{T}}(\dot x(t)) + \mathcal{H}_{\mathcal{T}}(\dot x(t))^\top P \mathcal{H}_{\mathcal{T}}(x(t)).
 	\label{lp}
 \end{multline}
 Then, using the solution $P^*$ of (\ref{optproblqr}), determine $\Gamma$ by solving the set of linear equations
 \begin{equation}
 	\left[ \begin{array}{c} \mathcal{H}_{\mathcal{T}}(x(t)) \\ \mathcal{L}(P^*) \end{array} \right] \Gamma = \left[ \begin{array}{c} I \\ 0 \end{array} \right].
 	\label{gammasol}
 \end{equation}
 In Lemma \ref{lemlqr} and Theorem \ref{thmlqr} below, we show that with the resulting value of $\Gamma$ from (\ref{optproblqr}) and (\ref{gammasol}), the solution to the CT LQR problem is given by\footnote{Instead of the identity matrix, any nonsingular matrix can be used in the first block row of the right-hand side of (\ref{gammasol}). In that case, the optimal LQR solution is given by $K^* = -\mathcal{H}_{\mathcal{T}}(u) \Gamma \left( \mathcal{H}_{\mathcal{T}}(x(t)) \Gamma \right)^{-1}$ (compare with \eqref{constabu}-\eqref{constabx}).} $K^* = -\mathcal{H}_{\mathcal{T}}(u) \Gamma$ .
 
 \begin{lem}
 	\label{lemlqr}
 	Consider the controllable system (\ref{ctsys}) and the cost (\ref{cost}) with $(A,Q^{1/2})$ detectable. Let $u$ be a PCPE input of order $n+1$, and let $x$, $\dot x$ be the noise-free state and state derivative measurements of (\ref{ctsys}). Then problem (\ref{optproblqr}) is feasible, its solution $P^*$ is unique, and $P^*$ satisfies the ARE (\ref{riccati}).
 \end{lem}
  \begin{pf}
 	Define $\mathcal{L}(P)$ as in (\ref{lp}), and the matrices
 	\begin{equation}
 		\bar{\mathcal{H}} := \left[ \begin{array}{c} \mathcal{H}_{\mathcal{T}}(x(t)) \\ \mathcal{H}_{\mathcal{T}}(u) \end{array} \right], \; \bar Q := \left[ \begin{array}{cc} Q + PA + A^\top P & PB \\ B^\top P & R \end{array} \right].
 		\label{hbar}
 	\end{equation}
 	Using (\ref{hdyn}), rewrite the LMI $\mathcal{L}(P) \succeq 0$ as $\bar{\mathcal{H}}^\top \bar Q \bar{\mathcal{H}} \succeq 0$. Since (\ref{pecond}) holds by persistence of excitation, this inequality holds if and only if $\bar Q \succeq 0$. Using the Schur complement and the fact that $R \succ 0$, we notice that $\bar Q \succeq 0$ if and only if $Q + PA + A^\top P - PBR^{-1}B^\top P \succeq 0$. Thus, one feasible solution to this problem is the solution $P^* \succ 0$ of the ARE (\ref{riccati}). By Lemma \ref{lemlqr}, $P^*-P \succeq 0$ for any other feasible solution $P$ of (\ref{optproblqr}), and therefore $\trace(P^*) \geq \trace(P)$. The proof is completed by showing that, if $P \neq P^*$, then $\trace(P^*) > \trace(P)$. This can be seen by noticing that, unless $P = P^*$, $\trace(P^* - P)=0$ implies the presence of both positive and negative eigenvalues in $P^*-P$, contradicting $P^*-P \succeq 0$. $\square$
 \end{pf}
 
 \begin{thm}
 	\label{thmlqr}
 	Let the conditions in Lemma \ref{lemlqr} hold. Moreover, let $\Gamma$ be computed as in (\ref{gammasol}), where $P^*$ is the solution to the convex optimization problem (\ref{optproblqr}). Then, the matrix $K = -\mathcal{H}_{\mathcal{T}}(u) \Gamma$ corresponds to the unique solution of the LQR problem defined by the cost (\ref{cost}).
 \end{thm}
 \begin{pf}
 	From (\ref{gammasol}), we have that $\Gamma^\top \mathcal{L}(P^*) \Gamma = 0$ and $\mathcal{H}_{\mathcal{T}}(x(t)) \Gamma = I$. This second equation, together with $K = -\mathcal{H}_{\mathcal{T}}(u) \Gamma$, implies that $\mathcal{H}_{\mathcal{T}}(\dot x(t)) \Gamma = A-BK$. This can be seen similarly as in the proof of Theorem~\ref{thmstab} (see (\ref{constabu}), (\ref{constabx}) and (\ref{clgamma})). Substituting these expressions in $\Gamma^\top \mathcal{L}(P^*) \Gamma = 0$, with $\mathcal{L}(P^*)$ as in (\ref{lp}), yields
 	\begin{equation}
 		Q + K^\top R K + P^*(A-BK) + (A-BK)^\top P^* = 0.
 		\label{bellmaneqk}
 	\end{equation}
 	By Lemma \ref{lemlqr}, $P^*$ also satisfies (\ref{bellmaneq}), where $K^*$ is the optimal LQR solution. Subtracting (\ref{bellmaneq}) from (\ref{bellmaneqk}), we obtain $K^\top R K - K^{*\top} R K^* - P^* B (K - K^*) - (K-K^*)^\top B^\top P^* = 0$. Finally, using the fact that $B^\top P^* = RK^*$, we get $(K - K^*) R (K - K^*)=0$, which implies $K = K^*$. $\square$
 \end{pf}

\subsection{Data-based robust pole placement}
\label{secpp}

Our objective in this section is to design a data-based algorithm that places the poles of system (\ref{ctsys}) in desired locations. This goal is formalized as follows.

\begin{prob}
	\label{probpp}
	Let $\Lambda = \{ \lambda_1, \ldots, \lambda_n \}$ be a self-conjugate set of complex numbers. Using data measured from (\ref{ctsys}), determine a state-feedback matrix $K$ such that the poles of $A-BK$ are located at the positions specified by $\Lambda$.
\end{prob}

A model-based solution to the robust exact pole placement problem was described in \cite{SchmidPanNgu2014}. In the following, we present a data-based variant of this method. As in \cite{SchmidPanNgu2014}, we first present a parameterization of all the matrices $K$ that solve Problem \ref{probpp}. This is presented for noise-free data, and then we show how to select $K$ for robust pole placement in the case of noisy data. Thus, consider the set $\bar{\Lambda} = \{ \lambda_1, \ldots, \lambda_\nu \} \subseteq \Lambda$, such that the algebraic multiplicity of $\lambda_i$ is $\eta_i$, and $\eta_1+\cdots+\eta_\nu = n$. Moreover, for some scalar $s \geq 0$, let the values $\lambda_i$, $i \leq 2s$, be complex numbers while $\lambda_i$, $i \geq 2s+1$, are real. Finally, for all odd $i <2s$, $\lambda_{i+1} = \lambda_i^*$. 

For $\lambda_i \in \bar{\Lambda}$, fix a $t \in [0,T]$ and define the matrices
\begin{equation}
	S(\lambda_i) := \mathcal{H}_{\mathcal{T}}(\dot x(t)) - \lambda_i \mathcal{H}_{\mathcal{T}}(x(t)), \quad i=1,\ldots,\nu.
	\label{slambda}
\end{equation}
Let $\bar{\mathcal{N}}_i \in \mathbb{C}^{N \times \bar s_i}$, be a basis for the right null space of $S(\lambda_i)$. Moreover, for the given $t$ and the matrix $\bar{\mathcal{H}}$ in \eqref{hbar}, let $\mathcal{N}_i \in \mathbb{C}^{n+m \times s_i}$ be a set of linearly independent columns of the product $\bar{\mathcal{H}} \bar{\mathcal{N}}_i$, such that $\text{rank }\left( \mathcal{N}_i \right) = \text{rank } \left( \bar{\mathcal{H}} \bar{\mathcal{N}}_i \right)$.
With the matrices $\mathcal{N}_i$, define
\begin{equation}
	\mathcal{N} = \row \{ \mathcal{N}_i \}_{i=1}^\nu = \left[ \mathcal{N}_1 \quad \cdots \quad \mathcal{N}_\nu \right].
	\label{nbasis}
\end{equation}
The following theorem shows a parameterization of the set of matrices $K$ that solve Problem~\ref{probpp}.

\begin{thm}
	\label{thmpp}
	Consider the controllable system (\ref{ctsys}). Let $u$ be a PCPE input of order $n+1$, and let $x$, $\dot x$, be the corresponding state and state derivative measurements \eqref{noisymes} with $\varepsilon_1 \equiv \varepsilon_2 \equiv 0$. Using the data $(x,\dot x)$, compute the matrices $S(\lambda_i)$ in (\ref{slambda}) and construct $\mathcal{N}$ in (\ref{nbasis}). Let $G := \diag \{ G_i \}_{i=1}^\nu$, $G_i \in \mathbb{C}^{s_i \times \eta_i}$, be an arbitrary parameter matrix such that for all odd $i < 2s$, $G_i$ is complex with $G_{i+1} = G_i^*$, and for all $i>2s$, $G_i$ is real. Define the following matrices
	\begin{IEEEeqnarray}{c}
		M(G) = \mathcal{N} G, \label{ppmg}\\
		X(G) = M(G)_{(1:n)}, \\
		V(G) = \mathcal{R}(M(G))_{(1:n)}, \label{ppvg} \\
		W(G) = \mathcal{R}(M(G))_{(n+1:n+m)}.
	\end{IEEEeqnarray}
	For almost every choice of $G$, $X(G)$ has full row rank. Moreover, the set of all gain matrices $K$ that solve Problem~\ref{probpp} is given by $K(G) = W(G) V(G)^{-1}$, 
	where $G$ is such that $X(G)$ has full row rank.
\end{thm}
\begin{pf}
	First, we show that each matrix $\mathcal{N}_i$ in (\ref{nbasis}) is a basis for the null space of the model-based matrix $\left[ A-\lambda_i I \quad B \right]$, $\lambda_i \in \bar{\Lambda}$. Recall that $\mathcal{N}_i$ is constructed with a set of linearly independent columns of the product $\bar{\mathcal{H}} \bar{\mathcal{N}}_i$ with $\bar{\mathcal{H}}$ as in \eqref{hbar}. Moreover, using (\ref{hdyn}), notice that
	\begin{equation*}
		\left[ A-\lambda_i I \quad B \right] \bar{\mathcal{H}} \bar{\mathcal{N}}_i = \Bigl( \mathcal{H}_{\mathcal{T}}(\dot x(t)) - \lambda_i \mathcal{H}_{\mathcal{T}}(x(t)) \Bigl) \bar{\mathcal{N}}_i = 0,
	\end{equation*}
	where the last equality follows from the definition of $\bar{\mathcal{N}}_i$ as a basis for the null space of (\ref{slambda}). This implies that $\left[ A-\lambda_i I \quad B \right] \mathcal{N}_i = 0$. Furthermore, consider any vector $v$ such that $\left[ A-\lambda_i I \quad B \right] v = 0$. Since the conditions in Lemma~\ref{lempe} are satisfied, (\ref{pecond}) holds and we can always find a vector $w$ such that $\bar{\mathcal{H}} w = v$. Therefore,
	\begin{equation*}
		0 = \left[ A-\lambda_i I \quad B \right] v = \Bigl( \mathcal{H}_{\mathcal{T}}(\dot x(t)) - \lambda_i \mathcal{H}_{\mathcal{T}}(x(t)) \Bigl) w,
	\end{equation*}	
	and, by definition of $\bar{\mathcal{N}}_i$, there is a vector $\bar w$ such that we can write $w = \bar{\mathcal{N}}_i \bar w$. The fact that $\left[ A-\lambda_i I \quad B \right] v = \left[ A-\lambda_i I \quad B \right] \bar{\mathcal{H}} \bar{\mathcal{N}}_i \bar w$ shows that $v$ can always be expressed in terms of $\mathcal{N}_i$, implying that $\mathcal{N}_i$ is a basis for the null space of $\left[ A-\lambda_i I \quad B \right]$.
	
	From this point, the rest of the proof follows as in the proof of \cite[Proposition 2.1]{SchmidPanNgu2014}. $\square$
\end{pf}

Theorem~\ref{thmpp} shows that the solution $K(G)$ of Problem~\ref{probpp} is parameterized by a matrix $G$, which provides degrees of freedom to design the controller. This fact has been used in different pole placement design methods to provide robustness against model uncertainties \citep{SchmidPanNgu2014,Varga2000}. We now discuss how this procedure is applicable in the case of noisy measurements as in \eqref{noisymes}. If Assumption~\ref{assnpe} holds, there exist unique matrices $(\hat A,\hat B)$ such that $\mathcal{H}_{\mathcal{T}}(\dot x(t)) = \hat A \mathcal{H}_{\mathcal{T}}(x(t)) + \hat B \mathcal{H}_{\mathcal{T}}(u) = [ \hat A \;\; \hat B ] \begin{bmatrix} \mathcal{H}_{\mathcal{T}}(x(t)) \\ \mathcal{H}_{\mathcal{T}}(u) \end{bmatrix}$. The difference between $(\hat A,\hat B)$ and $(A,B)$ depends on the measurement noise and, hence, using the data $x,\, \dot x$ with $\varepsilon_1,\varepsilon_2 \neq 0$ is analogous to the presence of model uncertainties. Following the results in \cite{SchmidPanNgu2014}, a computationally efficient method to robustly place the poles in their desired locations is to solve the optimization problem
\begin{equation}
	\underset{G}{\text{minimize}} \quad \| V(G) \|_F + \| V(G)^{-1} \|_F
	\label{optprobpp}
\end{equation}
subject to the conditions in Theorem \ref{thmpp}. Methods to solve (\ref{optprobpp}) have been developed in the literature \citep{SchmidPanNgu2014}.

\section{Simulation examples}
\label{secsim}

\subsection{Trajectory-reference control}
\label{secsimtr}

Our solution to the trajectory-reference control problem (see (\ref{optprobbc}) and (\ref{optprobbc2})) is tested on the linearized model of an aircraft. The model is given by (\ref{ctsys}), with $A$ and $B$ as in \cite{LuoLan1995}.
Suppose that we have available a single desired trajectory for the states of this system, such that $\Xi(t)$ and $\dot \Xi(t)$ in (\ref{exptraj})-(\ref{expder}) have a single column, e.g., $\Xi(t) = \xi^1(t)$. This reference trajectory was generated by a system of the form $\dot \xi^1(t) = \bar A \xi^1(t)$, where
\begin{equation*}
	\bar A =\left[ \begin{array}{cccc} -0.5254 &   0.0399 &  -1.4516 &   0.1061 \\	-1.8232 &  -2.4526  &  1.8725 &  -0.6407 \\	3.1222 &  -2.4746 &  -3.3309 &  -1.3357 \\	0.0046 &   1.3289  &  0.0157 &   0.0490 \end{array} \right].
\end{equation*}
These trajectories are such that they cannot be exactly followed by the states of the aircraft (\ref{ctsys}).

\begin{figure}
\begin{center}
\includegraphics[height=5cm,trim={0.4cm 0.6cm 1cm 0.3cm},clip]{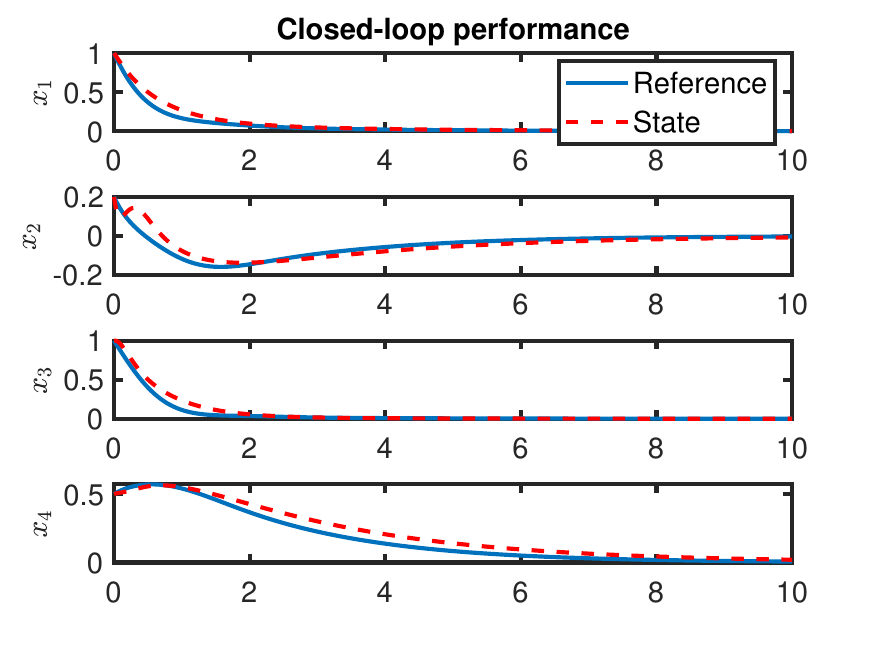}
\caption{A trajectory (blue) for each state is given as reference for problem (\ref{optprobbc}). The states of the system (dashed red) closely follow the desired references.}  
\label{figtr}
\end{center}                                 
\end{figure}

Measurement noise as in \eqref{noisymes} is generated as a uniformly distributed random signal such that $| \varepsilon_{i,j} | \leq 10^{-3}$ for each element $j$ of the vectors $\varepsilon_i$, $i=\{1,2\}$.  The output $\bar K$ of (\ref{optprobbc}) using these data is given by
\begin{equation*}
	\bar K =\left[ \begin{array}{cccc} 6.8377  &  0.5152  &  1.2275  &  0.6828 \\	-57.8280 &  -5.3991 &  -8.1967 &  -3.2920 \end{array} \right].
\end{equation*}
Since $\bar K$ is stabilizing, the problem (\ref{optprobbc2}) does not modify it, and $K = \bar K$ is the outcome of our method. When the initial states of the system coincide with $\xi^1(0)$, the closed-loop trajectories shown by red, dashed lines in Figure~\ref{figtr} are obtained. It can be seen that the reference trajectories are closely followed.

\subsection{Data-based optimal control}

In this subsection, we test the effect of noisy measurements in the procedure described in Section \ref{secoc} to solve the data-based LQR problem. From systems as in \eqref{ctsys}, we take measurements \eqref{noisymes} with uniformly distributed random signals $\varepsilon_1$, $\varepsilon_2$. Each element $j$ of the vectors $\varepsilon_i$, $i\in \{1,2\}$ is bounded as $| \varepsilon_{i,j} | \leq \bar \varepsilon$ for different values of $\bar \varepsilon$. 100 experiments with random stable systems \eqref{ctsys} with $n=4$ and $m=2$ are performed for each value of $\bar \varepsilon$.

In every case, the LQR problem with cost \eqref{cost} is solved for $Q=I_n$ and $R=2I_m$. After each experiment, we compare the obtained solution $K$ in Theorem~\ref{thmlqr} with the optimal LQR solution $K^*$. Table \ref{tab2} displays the average error for each value of $\bar \varepsilon$, computed as $\frac{1}{100} \sum_{i=1}^{100} \| K_i - K_i^* \|$. The obtained results show that the proposed method provides accurate results due to inherent robustness properties against measurement noise with small magnitudes.

\begin{table}
	\centering
	\begin{tabular}{|c|c|} 
		\hline
		$\bar \varepsilon$ & Average error \\[0.5ex]
		\hline
		$10^{-4}$ & $0.0096$ \\ 
		$10^{-3}$ & $0.0365$ \\
		$10^{-2}$ & $0.1497$ \\ [1ex]
		\hline
	\end{tabular}
	\vspace{4pt}
	\caption{Average error after 100 experiments with random systems using the results of Section \ref{secoc} with measurement noise.}
	\label{tab2}
\end{table}

\subsection{Robust pole placement}
\label{secsimpp}

In \cite{ByersNas1989}, the performance of a (model-based) robust pole placement algorithm is tested on eleven different systems, which have become benchmark examples in the literature \citep{SchmidPanNgu2014}. In this subsection, we apply the data-based method in Section~\ref{secpp} to these systems. Again, uniformly distributed noise of different magnitudes is added to the state data as in \eqref{noisymes}, where each element $j$ of $\varepsilon_i$, $i \in \{1,2\}$, is bounded as $| \varepsilon_{i,j} | \leq \bar \varepsilon$. Two different magnitudes of noise are considered, with $\bar \varepsilon = 10^{-3}$ and $\bar \varepsilon = 10^{-2}$. As a test for the accuracy of the obtained closed-loop eigenvalues, we compute $e = \sum_{i=1}^n | \lambda_i - \lambda_i^* |$, where $\lambda_i$ are the obtained eigenvalues, $\lambda_i^*$ are the desired eigenvalues, and the index $i$ sorts the eigenvalues in order of magnitude. To solve the optimization problem (\ref{optprobpp}), we used the Matlab toolbox kindly provided to us by the authors of \cite{SchmidPanNgu2014}; their (model-based) \textit{span} toolbox was modified by us to execute our data-based procedure.

Since the accuracy of the proposed method varies according to the PE data collected, we perform 100 experiments per example and average the errors $e$. In each experiment, the vectors $\mu$ in Definition \ref{defctpe} are selected randomly from a uniform distribution $\mathcal{U}(-5,5)^m$. Similarly, the initial state of the PE data is selected from $\mathcal{U}(-5,5)^n$. For comparison, the same noisy data is used for the data-based pole placement method in \cite{Bianchin2023}. The simulation results are shown in Table \ref{tab1}, where the acronym RPP refers to our robust pole placement procedure, while PP is used for the method in \cite{Bianchin2023}. The table shows that, as expected, the RPP algorithm outperforms the PP one. 

\begin{table}
	\centering
	\begin{tabular}{|c|c c|c c|} 
		\hline
		\multirow{2}{1.2cm}{\centering Example number} & \multicolumn{2}{c|}{Avg. error, $\bar \varepsilon = 10^{-3}$} & \multicolumn{2}{c|}{Avg. error, $\bar \varepsilon = 10^{-2}$} \\
		& RPP & PP & RPP & PP \\[0.5ex]
		\hline
		$1$ & $0.0053$ & $0.0832$ & $0.0240$ & $0.0553$ \\ 
		$2$ & $0.2512$ & $0.8299$ & $1.6981$ & $2.6842$ \\
		$3$ & $0.0395$ & $0.1010$ & $0.3894$ & $0.8933$ \\
		$4$ & $0.0482$ & $0.2125$ & $0.4614$ & $1.0039$ \\
		$5$ & $0.0053$ & $0.0301$ & $0.0420$ & $0.0653$ \\
		$6$ & $0.1625$ & $0.4237$ & $1.2576$ & $11.2209$ \\
		$7$ & $0.0135$ & $0.1836$ & $0.1070$ & $0.8459$ \\
		$8$ & $0.0134$ & $0.0305$ & $0.0889$ & $0.2173$ \\
		$9$ & $0.1821$ & $0.3501$ & $1.9451$ & $2.4716$ \\
		$10$ & $0.0061$ & $0.0205$ & $0.0365$ & $0.1211$ \\
		$11$ & $27.5198$ & $216.1408$ & $32.1955$ & $127.6611$ \\ [1ex]
		\hline
	\end{tabular}
	\vspace{4pt}
	\caption{Average error $ \sum_i | \lambda_i - \lambda_i^* |$ comparison between the method in Section \ref{secpp} (RPP), and the method in \protect\cite{Bianchin2023} (PP). The example numbers are from \cite{ByersNas1989}.}
	\label{tab1}
\end{table}

\section{Conclusion}
\label{secconc}

In this paper, we presented the solution to three different data-based control problems for continuous-time systems, where specific closed-loop performances are required. The first problem is a trajectory-reference control problem, where desired closed-loop trajectories are given and a controller is designed to make the system follow them as closely as possible. Then, we solved the data-based LQR problem for continuous-time systems. The final problem is a robust pole placement procedure which, different from other results in the literature, addresses the \emph{exact} pole placement problem while considering robustness against the effect of noisy measurements. The solution to these problems was facilitated by leveraging the data-based continuous-time system representation proposed in \cite{LopezMuCDC2022}. Following analogous steps, the reader can note that all of these problems can be formulated and solved for discrete-time systems as well. An important direction for future work is the development of control protocols for (classes of) nonlinear CT systems, similarly as it has been done in the DT case.

\begin{ack}                               
	This work has received funding from the European Research Council (ERC) under the European Union’s Horizon 2020 research and innovation programme (grant agreement No 948679).
\end{ack}

{\scriptsize \bibliographystyle{apalike}
\bibliography{dbcontrol_refs}}

\end{document}